\documentclass[a4]{article}
\usepackage{amssymb}

\newcommand{\Complex}{\mathbb{C}}

\usepackage{epsfig}

\usepackage{epstopdf}
\begin{document}



\title{Numerical treatment of interfaces in Quantum Mechanics}
\author{Oscar Reula~\footnote{FaMAF-UNC, IFEG-CONICET, Ciudad Universitaria, 5000, C\'ordoba, Argentina}}
\date{March 28, 2011}

\maketitle

\begin{abstract}
{In this article we develop a numerical scheme to deal with interfaces between touching numerical grids when solving 
Schr\"o{}dinger equation. In order to pass the information among grids we use the values of the fields only at the contact
point between them. Surprisingly we obtain a convergent methods which is third order accurate with respect to the spatial
resolution. In test cases, at the minimal resolution needed to describe correctly the waves, the error of this approximation 
is similar to that of a homogeneous (centered differences everywhere) 
scheme with three points stencil, that is a sixth order finite difference operator. 
The semi-discrete approximation preserves the norm and uses standard finite difference operators satisfying summation by parts. 
For the time integrator we use a semi-implicit IMEX Runge Kutta method.}   

\end{abstract}





\section{Introduction}

For some years now there has been numerical techniques to deal with interfaces (boundaries between numerical grids)
when solving hyperbolic or parabolic equations. 
Some of them use interpolation between regions of overlap, while others use penalties which modify the system at boundary grid points including information from the same space points at other grids \cite{Carpenter1999341}. 
This last method is preferable in many situations for it has very nice properties, the more interesting one is the fact that it is constructed
so that the resulting semi-discrete system preserves the corresponding continuum energy estimate of the corresponding constant coefficient linear systems. Thus we can ensure that, at least for linear, constant coefficients system, the scheme is stable.

Another important property this technique has arises when dealing with parallel codes using, say MPI. 
In that case, data near the boundaries is needed to be passed along different MPI process (usually a grid or part of it), 
each one of them usually running at different hosts, and so it is important to pass as little information among them as possible. 
For these cases this method is optimal, since only the boundary points need to be shared among  grids the amount of data among processes is minimal and does not depend on the precision of the numerical scheme.
This property is also important since otherwise interpolation is needed when the geometry of the grids do not coincide beyond the 
shared boundary points.

In this paper we show that the penalty method not only works for hyperbolic and parabolic systems but also can be extended to the Schr\"odinger equation, having similar properties as the ones mentioned above.

In the next section we derive the boundary terms needed to be added to the equations for the method to work, is this case there are two different types of  terms. Only one of them depends only on the values of the fields at the same grid, its presence is needed to cancel the usual boundary term from the elliptic part of the operator, the other term is really an interaction between the fields at both grids, it is a penalty only in the sense that depends on the 
difference of fields on both sides of the grids, but contrary to the hyperbolic and parabolic cases, where the penalties introduced a large negative eigenvalue to the system which would bring down to zero the difference on both sides of the grid, in this case the contribution is a large purely imaginary
eigenvalue which keeps the $l^{2}$ norm constant, but nevertheless does the allow the waves to pass through the interface from one grid to the other.

In the last section we show some numerical results using this method. We compare the results of evolving a one dimensional system in a 
circle, first using periodic boundary conditions in a single grid (that is a homogeneous scheme using centered difference operators) and then using the interface scheme between the first and final point of the grid. 
We show convergence of the scheme compatible with the discretization method used and discuss the quality of the approximation. 


\section{Numerical scheme}

We assume, for simplicity, we are in one dimension, the generalization to more dimensions is trivial.
So we consider a complex field $\Phi(x,t): S^{1} \times \Re \to \Complex$ satisfying Schr\"odinger equation:

\[
i\partial_{t} \Phi = \Delta \Phi + V(x,t)\Phi
\]
and assume initial data is given at $t=0$, $\Phi(x,0) = \Phi_{0}(x)$, such that it is sufficiently smooth.
The potential $V(x,t)$ is assumed to be smooth and compact support in the space direction. 
Standard theorems guaranty the existence of a solution in the $L^{2}$ norm. Since the potential is irrelevant for
interface considerations we just drop it from now on.

In order to solve numerically this problem we take an evenly spaced grid and set an interface at $x=0$ that 
will connect one grid end with the other, resulting in a circle of length one. We call the discrete solution by the vector, 
$\{\Psi_{j}\}$, $j=0\dots N$ corresponding to points $x_{j} = dx*j$ with $dx := \frac{1}{N}$ so that the last point coincides 
with the first one.  

We introduce the discrete $l^{2}$ norm as usual, 
\[
\label{l2_norm}
<\Psi,\Phi> := dx \sum_{i=0}^{N} \sigma_{j} \bar{\Psi}_{j} \Phi_{j}
\]
where $\{\sigma_{j}\}$ are a set of real valued weights that depend on the finite difference operators under consideration and $dx$ is the interspace
between neighboring grid points.

The semi-discrete system we want to solve at all points, except at the boundary is:

\[
\partial_{t} \Psi_{j} = -i (D^{2}\Psi)_{j} \;\;\;\;\;\; j = 0 \ldots N
\]
where $D$ is any finite difference operator approximating the derivative operator to some order $q>=1$ which satisfies the summation by parts
property (SBP from now on) \cite{Kreiss-Scherer:1977,Strand199447,gko1995}. That is, it satisfies,

\[
<\Psi,D\Phi> + <D \Psi,\Phi> :=  \bar{\Psi}_{N}\Phi_{N} - \bar{\Psi}_{0}\Phi_{0}
\]
If we could proof that this linear ODE system has eigenvalues with no positive real part and a complete set of eigenvectors, then there are many discrete time integrators, which would give a stable numerical evolution to the whole system (for a more detailed description of the theory see for instance \cite{gko1995}). 
A way to check those conditions is to find a norm which is preserved or decreases in time. This is the procedure we shall use to device our scheme. 
In particular we shall later use either traditional explicit Runge Kutta fourth order operators or new IMEX ones which are mixtures of implicit and explicit Runge Kutta methods.

It is clear then that if we use this scheme, we would get,

\[
\frac{\partial}{\partial t}<\Psi,\Psi>  := -i \bar{\Psi}_{N} D\Psi_{N} + i \bar{\Psi}_{0} D\Psi_{0} + c.c. 
\]
where $c.c.$ means complex conjugate terms. Since these contributions come from each side, in order to preserve this norm along evolution we need to cancel these boundary terms. Contrary to the hyperbolic and parabolic cases, it doesn't seem possible to do that by introducing on each side terms proportional to the difference on the fields and their normal derivatives at each boundary.  Thus, we introduce our first modification to the scheme by adding terms at the boundary as follows,

\[
\partial_{t} \Psi_{j} = -i (D^{2}\Psi)_{j} + i \frac{1}{dx \sigma_{0}}\delta_{j0} D\Psi_{0}  - i \frac{1}{dx \sigma_{N}}\delta_{jN} D\Psi_{N} 
\]
with this modification the boundary terms cancel and so the total norm remains constant, but there is no interaction among both sides of the interface, the solution we would get would just bounce back at the boundary (the probability is conserved and can not possibly influence the point at the other side for there is no interaction among them so has to bounce back). But eliminating the boundary term means that we now can concentrate on adding terms which, while preserving the norm, would introduce an interaction at the ends of the grid in such a way as to let the wave pass through the interface.
We must introduce a term which couples both sides, namely a penalty term which forces the values to both extremes to coincide. 
The simplest one which satisfies this property is:

\[
\partial_{t} \Psi_{j} = -i (D^{2}\Psi)_{j} + i \frac{1}{dx \sigma_{0}}\delta_{j0} D\Psi_{0}  - i \frac{1}{dx \sigma_{N}}\delta_{jN} D\Psi_{N}   
                               - iL(\Psi_{0}-\Psi_{N})(\delta_{j0}  - \delta_{jN})  
\]
where $L$, which we call the {\sl interaction factor}, is a real constant to be chosen as big as possible, in order to make the interaction as strong as possible while keeping the $l^{2}$ norm constant.
We could also contemplate a complex $L$, the addition of a negative imaginary part would make the discretization more stable, but a corresponding loss on the $l^{2}$ norm would occur, that is, in physical terms we would loose some probability of finding the particle each time it passes by the interface. 
When we tried it on the tests no improvement on accuracy was found over a very wide range of values. On the other hand, since we do not need any dissipation to improve accuracy so we do not use it here. 
The limitation on upper value of the real part of $L$ comes from the fact that a too large a value would  make the system unstable by making a too big contribution to the eigenvalues along the imaginary axis, making explicit time integration schemes to fall outside their stability region, or making the time step prohibive small. For explicit schemes the value of $L$ should not be bigger than  $L = \frac{1}{\sigma_{0}dx^{2}}$, so it contributes to the CFL factor as much as the principal part. This turns out not to be good enough, giving unacceptable large errors in the form of bounces at the interface for a resolution which describes appropriately the waves of the test.
Thus we had to use bigger factors and so to resort to a semi-implicit method which could free us from the CFL limitation.

The above system of ordinary differential equations was evolved using Runge Kutta type methods, first the usual fourth order one, and
then an IMEX method \cite{Ascher1997151, springerlink:10.1007/s10915-004-4636-4}, in particular the one called
IMEX-SSP3(4,3,3) L-stable scheme  in \cite{springerlink:10.1007/s10915-004-4636-4}. 
We report on the findings in the next section.

\section{Test}

We choose as initial data the following function:

\[
\Phi(x) := \exp^{\frac{-(x-1)^{2}}{20}}\exp^{\Im \pi 100 x}
\]
We run a base configuration with $N=2000$ points on a circle of length 2. The number of points and the order of the finite difference operators employed guarantee a good enough resolution for such a high frequency\footnote{Here we aim at an accuracy of about $10^{3}$ for 10 periods. We note that with these number of points and second order accurate centered difference operators the error is larger than this even for one period.}. 
We first evolved the solution up to $T=0.004$, at which the solution has moved to the left and the pulse has passed completely the interface, located at $x=0$, for a distance large enough as to not occupy the place a bouncing pulse would do.
In most of the runs we use a very accurate finite difference operator of order eight in the interior and order four at points in the boundary satisfying SBP\cite{Lehner:2005bz}. The choice of that operator was made to preserve the correct phase of the solution on long time runs, and to be able to probe the convergence error contribution of the interaction term with as smaller interference from the error of the other discretizations as possible.
For comparison we made runs using periodic boundary conditions with $8^{th}$ order centered difference operators with $N=2000$, $4000$ and $8000$ points, this last one is used as the reference solution agains which we compare all our runs. 
The numerical integrator in these cases was the fourth order Runge Kutta one.

\subsection{Traditional Runge Kutta Scheme}

\begin{figure}[htbp]
\begin{center}
\includegraphics[width=3.5in]{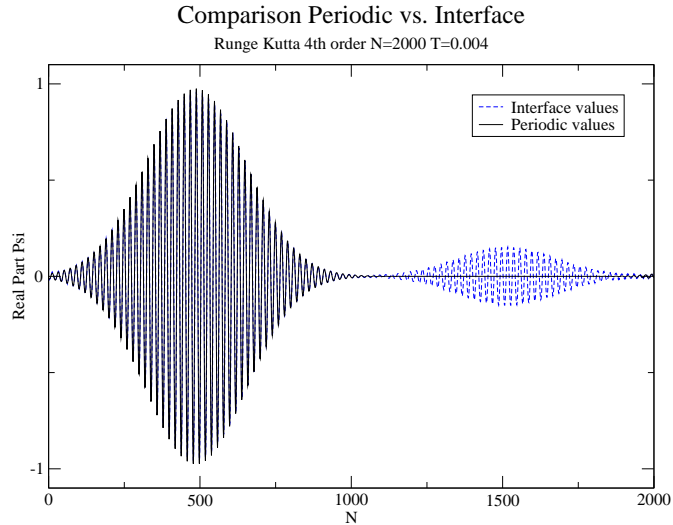}
\caption{{\bf Comparison between periodic and interface runs, using traditional $4^{th}$ order Runge Kutta method. Here $L=1000*dx^{-2}$.}}
\label{Comparison-RK}
\end{center}
\end{figure}

With the traditional Runge Kutta schemes one gets a converging scheme but the error is disappointingly large. This is due to the fact that we used
an interaction factor of $L = \frac{1}{\sigma_{0}dx^{2}}$. In the plot below, Figure \ref{Comparison-RK}, we show both the periodic and the interface approximations. 
The extra bump to the right is the bounce of a fraction of the solution at the interface. The part of the solution which goes through the interface keeps the phase very well. 

It is possible to reduce the error into really small limits by enlarging the interaction factor but at the expense of loosing efficiency since
the time step becomes significantly smaller, in fact we observed the error falls to very small values for an interaction factor a thousand times bigger. 
To avoid that in the next subsection we consider semi-implicit methods.

\subsection{IMEX Scheme}

To avoid small time steps while allowing larger interaction factors semi-implicit methods are needed. 
We shall use here a method among those called IMEX, \cite{Ascher1997151, springerlink:10.1007/s10915-004-4636-4}, in particular the one called
IMEX-SSP3(4,3,3) L-stable scheme  in \cite{springerlink:10.1007/s10915-004-4636-4}.
These methods permit to explicitly solve stiff parts of the equations but keeping the other terms as usual in traditional Runge Kutta schemes.
Although these methods are designed for systems with large negative real part eigenvalues, they seem to perform well 
in this highly oscillatory (eigenvalues with large imaginary part) case.

\begin{figure}[htbp]
\begin{center}
\includegraphics[width=3.5in]{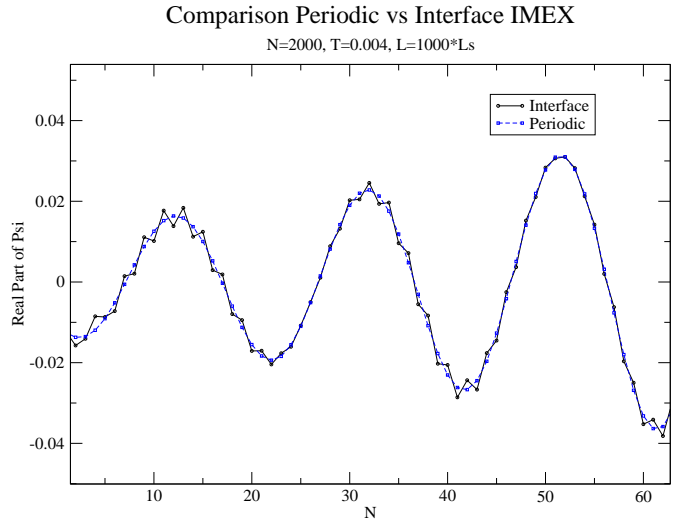}
\caption{{\bf Comparison between periodic and interface runs, using a Runge Kutta IMEX time integrator. Here $L=1000*dx^{-2}$.    }}
\label{ComparisonIMEX}
\end{center}
\end{figure}

Below, Figure \ref{ComparisonIMEX}, shows a comparison as the above but using an IMEX method and an interaction factor thousand times larger. 
In this case the error is so small that we just have zoomed into a small region near the interface to show how the solutions follow each
other in phase. We observe that the solution going through the interface has a small higher frequency component, there in no noticeable pick at the interface point.

\subsection{Convergence}

\begin{figure}[htbp]
\begin{center}
\includegraphics[width=3.5in]{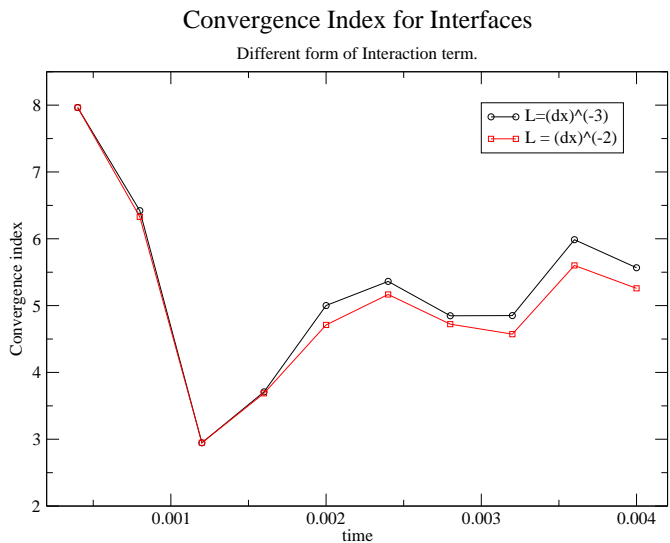}
\caption{{\bf Convergence index for two different ways of scaling the interaction factor with resolution.}}
\label{ConvergenceIMEX}
\end{center}
\end{figure}

In absence of the interaction term we expect the error to be of the form $e = f_{1}dt^{p} + f_{2}dx^{q}$, where $p$ depends on the time integrator used and $q$ on the space discretization of derivatives. In our case, we expect $p\geq 2$ for the IMEX algorithm (depending on the nature of the solution, in particular the size of the solution near the boundary, where the implicit part of the algorithm is used, in comparison with the size of the solution in the interior of the grid) and $q\geq 4$ since the derivatives used are four order accurate at the boundary and eight order accurate in the interior.
For stability reasons, the CFL condition on the explicit integrator, we need to scale $dt$ as $dx^{2}$ so we expect a convergence index of the order of
four. Any smaller convergence factor must result from the interface treatment.
We made two sets of convergence runs, one with the interaction factor scaling as $L = 10^{3} dx^{-2}$, and the other as  $L = 10^{3} dx^{-3}$.
From the plot below, Figure \ref{ConvergenceIMEX}, we see that the convergence factor has a minimum (when most of the solution is crossing the boundary) of about $q=3$, and 
that this minimum is the same for both scalings of the interaction factor.
Thus we see that the convergence we get with this method seems to be dominated by the interaction factor.

Convergence alone is not enough to guaranty that we are approaching the correct solution, this is so because our system is not appriory
consistent, that is, in principle the limit of our finite difference scheme does not need to coincide with the continuum equation because of the boundary terms which grow with resolution. Thus we really have to analyze convergence against the true solution, which we do in the next section.

\subsection{Accuracy}

\begin{figure}[htbp]
\begin{center}
\includegraphics[width=3.5in]{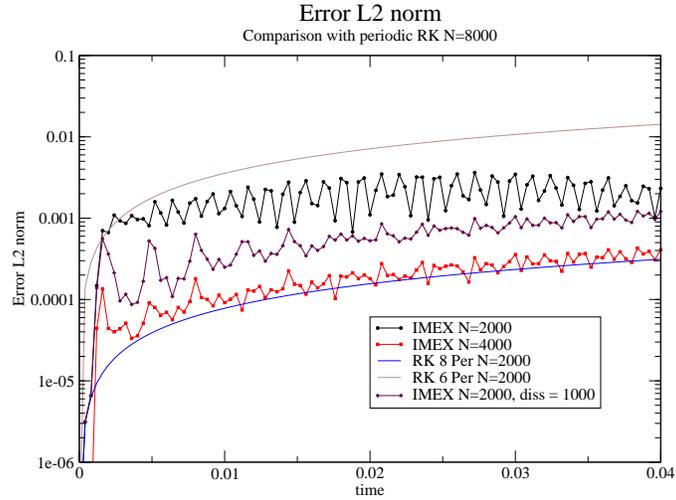}
\caption{{\bf $l^{2}$ norm of the error for several different runs. The rugged lines correspond to runs with the interface while the smother ones correspond to periodic ones.}}
\label{L2_comparison_IMEX_long}
\end{center}
\end{figure}

To see convergence to the right solution we computed a periodic solution (everywhere centered finite difference operators) with $N=8000$ and $T=0.04$ (ten times the time used before) and used it as reference solution to compute the error.
In the plot below, Figure \ref{L2_comparison_IMEX_long}, we can see that the interface approximation with $N=2000$ has an error smaller than a similar
run in the periodic setting with finite differences operators of $6^{th}$ order. So it is clear the scheme produces approximations converging to the 
correct solution. We point out that, even for the periodic setting, the phase coherence for lower order finite difference operators (second and fourth order accurate) is lost much earlier in time.
 
Although the total convergence is only third order the actual error the solution has is very small and does not seems to change the phase of the wave very much. 
We included also an interface approximation with $N=4000$ points which has a similar error to the periodic one with $8^{th}$ order finite difference operators and $N=2000$. 
For comparison we have included a run with Kreiss-Oliger dissipation (see \S \ref{Dissipation} below). We observe that for moderate values the 
accuracy improves a bit further, although at the expense of a faster decay of the $l^{2}$ norm.

\vspace{1cm}

\subsection{Norm preservation}

The present scheme is norm preserving at the semi-discrete approximation level. That implies that if a stable time integrator is used 
with a sufficiently small time step the norm should only be able to decrease at a rate given by the dissipation of that integrator. 
We find that the $l^{2}$ norm is preserved very well along evolution. 
See Figure \ref{L2_Norm_IMEX} below, where we have plotted the relative error in the norm for the low
resolution case, $N=2000$ and the IMEX integrator.  
It seems to grow a little bit when most of the solution is passing through the interface. 
This would seems to contradict what we has just stated. 
This paradox is due to the fact that the $l^{2}$ norm we are plotting corresponds to the trapezoidal rule approximation, ($\sigma_{0} = \sigma_{N}= \frac{1}{2}, \sigma_{i}=1, i=1\ldots N-1$ in \ref{l2_norm}) which is the norm associated with the second order accurate finite difference approximation but not to the one associated with the $8^{th}$ order one we are using. 
In fact, when the solution is several grid points away from the interface, where all the norms for the different finite difference operators coincide, we do indeed get a smaller value for it. 
When running the code with the lower second order finite difference operators, the norm indeed slowly and monotonically decreases as the theory predicts, but the solutions is almost immediately out of phase with respect to the reference one.  

\vspace{1cm}
\begin{figure}[htbp]
\begin{center}
\includegraphics[width=3.5in]{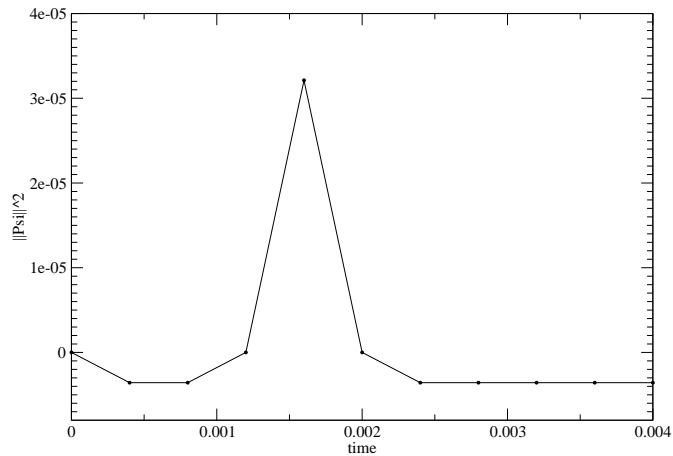}
\caption{{\bf Behavior of the relative error of the $l^{2}$ (trapezoidal rule) norm on a short run. The pick appears when the solution is at the boundary. It is only an effect of the use norm of a norm which is not exactly the corresponding one for the finite difference operator used.}}
\label{L2_Norm_IMEX}
\end{center}
\end{figure}

\begin{figure}[htbp]
\begin{center}
\includegraphics[width=3.5in]{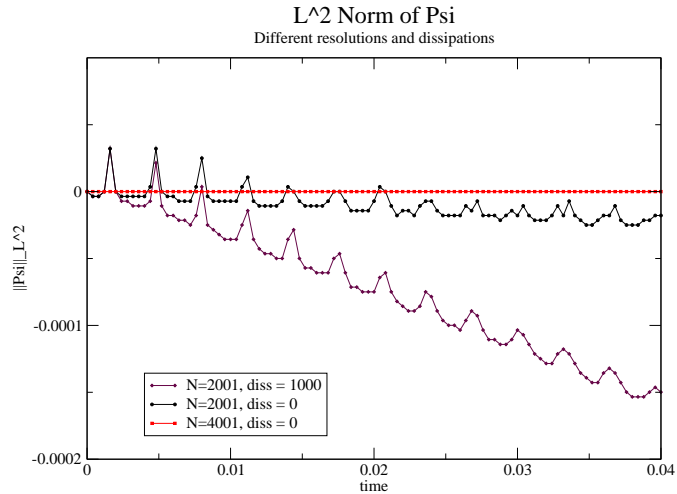}
\caption{{\bf Behavior of the  relative error of the $l^{2}$ norm for a longer run in order to see the decay due to the time integrator used.}}
\label{L2_Norm_IMEX_LONG}
\end{center}
\end{figure}

The last plot, Figure \ref{L2_Norm_IMEX_LONG}, shows the behavior of the norm on longer runs, ten times the previous ones. At the lowest resolution we
see the expected decay with the periodic bumps due to the use of the wrong approximation to the norm. A higher resolution run is also shown illustrating that in this case the decay is negligible. We have also included a run with Kreiss-Oliger dissipation, (see \S \ref{Dissipation} below).

\subsection{Dissipation}
\label{Dissipation}

Much of the noise the interface interaction produces can be taken away by using Kreiss-Oliger  dissipation \cite{gko1995}, that is, by adding
to the equations a term proportional to a large power of the Laplacian operator times a factor which depends on the resolution in such a way as to make the error produced by adding this term to be of the same or smaller order as the order of the  rest of the terms in the approximation. 
In our case we use a term proportional to the discretized version of the Laplacian to the fourth order. One can see, running at the lowest resolution, that the solution smoothes out in a nice way, and indeed the error becomes smaller, (see Figure \ref{L2_comparison_IMEX_long}).  But this is at the expense of a faster decay of the $l ^{2}$ norm (see Figure \ref{L2_Norm_IMEX_LONG}). Depending on the purpose of the calculation one can choose whether or not to use dissipation.


\section{Conclusions}

We have shown that it is possible to implement an interface scheme of the ``Penalty'' type for the Schr\"odinger equation similar to the ones 
used for first order hyperbolic and parabolic equations. It shares with them similar properties, only data at points at the interface need to be passed between grids, and convergence is ensured for linear, constant coefficient, systems. Although the scheme seems to be third order accurate, the actual error we find in our test, at the lowest reasonable  resolution, seems to compare very well with sixth order homogeneous (centered finite difference operators) schemes. 
This is important for multi-block parallelizations for it implies one obtain the same quality for a solution only sharing $(1/3)^{n}$, $n$ being the space dimension, of the data one would need for a comparable (in accuracy) homogeneous scheme, that is, for low but reasonable resolutions, only passing one point at the boundary instead of three points a sixth order centered difference operator would need.

\section{Acknowledgementes}

We thanks Luis Lehner for discussions, and SeCyT-UNC, CONICET, FONCyT and the Partner Group grant of the Albert-Einstein Max Plack Institute for Gravitational Physics for financial support.

\bibliography{QM}

\end{document}